\documentclass[%
    reprint,
    superscriptaddress,
    longbibliography,
    amsmath,amssymb,
    balancelastpage,
    aip,apl]{revtex4-2}
\usepackage{graphicx}   
 \usepackage[none]{hyphenat}
\usepackage{siunitx}
\usepackage{ORCIDinREVTeX}

\usepackage[colorlinks=true,
    linkcolor=blue,     
    citecolor=blue,     
    filecolor=blue,     
    urlcolor=blue]{hyperref}
\allowdisplaybreaks

\newcommand{\figref}[1]{\hyperref[{#1}]{\textup{Fig.~\ref*{#1}}}}
\renewcommand{\eqref}[1]{\hyperref[{#1}]{\textup{(\ref*{#1})}}}
\newcommand{\barkappa}{\tilde{\kappa}}

\begin{document}
\title{Wireless Real-Time Capacitance Readout Based on Perturbed Nonlinear Parity-Time Symmetry}

\author{Ke Yin}
\orcid{0000-0002-8534-216X}
\affiliation{\mbox{School of Electrical Engineering, Xi'an Jiaotong University, Xi'an 710049, China}}

\author{Yuangen Huang}
\affiliation{\mbox{School of Electrical Engineering, Xi'an Jiaotong University, Xi'an 710049, China}}

\author{Chao Ma}
\orcid{0000-0003-3952-0516}
\affiliation{\mbox{School of Electronics, Peking University, Beijing 100871, China}}

\author{Xianglin Hao}
\orcid{0000-0001-7149-0113}
\affiliation{\mbox{School of Electrical Engineering, Xi'an Jiaotong University, Xi'an 710049, China}}

\author{Xiaoke Gao}
\affiliation{\mbox{School of Electrical Engineering, Xi'an Jiaotong University, Xi'an 710049, China}}

\author{Xikui Ma}
\affiliation{\mbox{School of Electrical Engineering, Xi'an Jiaotong University, Xi'an 710049, China}}

\author{Tianyu Dong}
\orcid{0000-0003-4816-0073}
\email[Author to whom correspondence should be addressed. Electronic mail: ]{tydong@mail.xjtu.edu.cn}
\affiliation{\mbox{School of Electrical Engineering, Xi'an Jiaotong University, Xi'an 710049, China}}

\date{March 31, 2022}

\begin{abstract}
  In this article, we report a vector-network-analyzer-free and real-time LC wireless capacitance readout system based on perturbed nonlinear parity-time (PT) symmetry. The system is composed of two inductively coupled reader-sensor parallel RLC resonators with gain and loss respectively. By searching for the real mode that requires the minimum saturation gain, the steady-state frequency evolution as a function of the sensor capacitance perturbation is analytically deduced. The proposed system can work in different modes by setting different perturbation point. In particular, at the exceptional point of PT symmetry, the system exhibits high sensitivity. Experimental demonstrations revealed the viability of the proposed readout mechanism by measuring the steady-state frequency of the reader resonator in response to the change of trimmer capacitor on the sensor side. Our findings could impact many emerging applications such as implantable medical device for health monitoring, parameter detection in harsh environment and sealed food packages, \emph{etc}.
\end{abstract}

\maketitle

\paragraph*{Introduction \label{sec:intro}}
The concept of parity-time (PT) symmetry is first presented in the context of quantum mechanics by C. M. Bender and S. Boettcher, which states that the Hamiltonian of a quantum system has not to be Hermitian but PT-symmetric to guarantee real energy levels of the system which are physically observable, extending conventional quantum mechanics to complex domain \cite{bender1998real,bender2002complex,bender2007making}. Although PT symmetry has its roots in quantum mechanics, extensive research have shown intriguing phenomena of non-Hermitian systems in various other fields, including optics \cite{makris2008beam,guo2009observation,ruter2010observation,zhu2016asymmetric}, electronics\cite{schindler2011experimental,lin2012experimental,schindler2012symmetric,bendern2013observation,assawaworrarit2017robust,choi2018observation,chen2018generalized,sakhdari2018ultrasensitive}, microwaves\cite{bittner2012pt,yu2020phase}, and acoustics\cite{zhu2014pt,fleury2015invisible,popa2014non}, to mention a few. Among all these fields, electronics provides an easily accessible platform to investigate all the characteristics of PT symmetry and has been extensively investigated \cite{schindler2011experimental,lin2012experimental,schindler2012symmetric,bendern2013observation,assawaworrarit2017robust,choi2018observation,chen2018generalized,sakhdari2018ultrasensitive}. Also, it paves the way towards the creations and applications of novel electronic systems and devices, including robust wireless power transmission (WPT) \cite{assawaworrarit2017robust} and wireless microsensor readout with enhanced sensitivity \cite{chen2018generalized,sakhdari2018ultrasensitive}, \emph{etc}.

In this article, we focus on one of the important applications of PT symmetry in electronics, \emph{viz.}, PT-symmetry-enhanced wireless sensor readout. Conventional LC passive microsensor readout is typically based on measuring the dip shift of the reflection spectrum from the readout coil connected to the vector network analyzer (VNA) \cite{duan2016wireless,boutry2019biodegradable,liang2019lc}, whose readout performance is unsatisfactory due to the poor sensitivity and resolution restricted by the small size of the microsensor. Aiming to this issue, PT symmetry provides a new strategy for improving the readout performance thanks to the eigenfrequency splitting phenomena nearby the exceptional point (EP) \cite{li2019sensing,sakhdari2018ultrasensitive,hajizadegan2019high,chen2018generalized}. In the previous readout systems \cite{chen2018generalized,sakhdari2018ultrasensitive}, a series-RLC active reader incorporating VNA together with an inductively coupled capacitive pressure microsensor constitutes a PT-symmetric gain-loss electronic dimer, by which both the resolution and sensitivity of the measured reflection spectrum can be enhanced. Nevertheless, a prerequisite that the reader capacitance must be equal to the sensor capacitance is required to guarantee PT symmetry of the entire system. Thus, the reader capacitor needs be tuned manually to follow the variation of the sensor capacitor, resulting in that such a proof-of-concept is not suitable for practical application. In a later work, the reader capacitor was replaced by a varactor \cite{zhou2020enhancing}. By cyclically scanning the bias voltage of the varactor, the reflection spectrum measured via VNA with the greatest Q factor corresponding to the PT-symmetric case can be extracted, which is time consuming thus cannot realize real-time sensing. Moreover, in either conventional \cite{duan2016wireless,boutry2019biodegradable,liang2019lc} and PT-symmetric \cite{chen2018generalized,sakhdari2018ultrasensitive,zhou2020enhancing} readout scheme, the readout circuit needs to be connected to VNA for reflection spectrum measurement, which is expensive and bulky thus not suitable for applications outside the laboratory.

Toward this end, we have demonstrated here that PT symmetry can enable a VNA-free readout by directly monitoring the working frequency of the readout circuit. Compared to previous works, the system functions in time domain thus avoiding frequency sweep on the reader side. Different from measuring the reflection spectrum when the dip frequency is extracted for each capacitance sensing point, the proposed readout mechanism can realize real-time sensing since the circuit works at the dip frequency \emph{per se} and changes simultaneously along with the capacitance variation of the sensor. Furthermore, the capacitor on the reader side does not need to be manually tuned to satisfy the PT-symmetry condition. Our work not only offers new design strategies for VNA-free LC wireless sensors but also provides a frequency chosen mechanism for PT-symmetry-enhanced WPT systems.

\begin{figure}[!ht]
  \centering
  \includegraphics{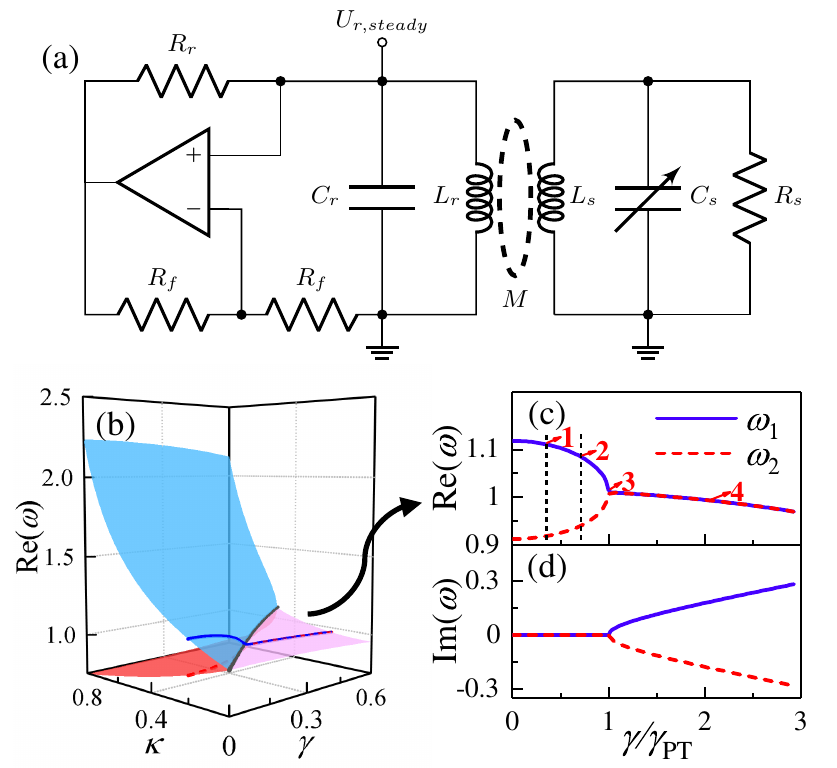}
  \caption{(Color online) (a) Parallel-parallel LC wireless capacitive sensing system with nonlinear gain. (b) Real part of the eigen-frequency evolution with respect to the gain/loss parameter $\gamma$ and coupling coefficient $\kappa$ when $\gamma = \gamma_s = g_r = \frac{1}{R_s} \sqrt{\frac{L_r}{C_r}}$ and $\eta=1$. (c) Real and (d) imaginary part of eigen-frequency evolution with respect to the normalized gain/loss parameter $\gamma/\gamma_\text{PT}$ when $\kappa=0.2$. The labeled points in (c) corresponds to different perturbation points for capacitance sensing.\label{fig:figure01} }
\end{figure}

\paragraph*{Sensing mechanism: Steady-state frequency vs. capacitance \label{sec:theory}}
\figref{fig:figure01}(a) shows the circuit schematics of the proposed wireless sensing system, which consists of two coupled parallel RLC resonators, one with nonlinear gain and the other with loss. The nonlinear gain originates from the operational amplifier (OP AMP) which is constructed as a non-inverting amplifier with closed-loop voltage gain $A_u = 2$ functioning as a negative impedance converter. With the resistor $R_r$ providing positive feedback, the circuit can realize a negative resistor $-R_r$ in the linear range of OP AMP. 

We aim to detect the varying of the capacitance $C_s$ wirelessly by virtue of the \emph{so-called} perturbed parity-time symmetry. By applying Kirchoff's laws, the system equations can be derived as two coupled second-order ordinary differential equations given by
\begin{subequations} \label{eq:system_equation_differential}
  \begin{align}
    \frac{\text{d}^2 U_r}{\text{d} \tau^2} - g_r(U_r) \frac{\text{d} U_r}{\text{d} \tau} + \frac{1}{\barkappa}U_r                   & = \frac{\kappa}{\barkappa}U_s,  \label{eq:system_equation_differential_a}      \\
    \frac{\text{d}^2 U_s}{\text{d} \tau^2} + \frac{\gamma_s}{\eta}\frac{\text{d} U_s}{\text{d} \tau} + \frac{1}{\eta\barkappa}U_s & = \frac{\kappa}{\eta\barkappa}U_r, \label{eq:system_equation_differential_b}
  \end{align}
\end{subequations}
where the system variables $U_r, U_s$ denote the voltages of the capacitor on the reader and sensor side, respectively; $\eta = C_s/C_r$ reflects the perturbation caused by capacitive sensor; $\kappa = M/\sqrt{L_r L_s} = M/L$ is the coupling coefficient; $\barkappa = 1 - \kappa^2$; $\tau = \omega_r t = t/\sqrt{L_r C_r}$ is the normalized time; $\gamma_s = \frac{1}{R_s} \sqrt{\frac{L_r}{C_r}}$ is the normalized loss parameter of the sensor resonator; $g_r(U_r)$ is the normalized nonlinear gain parameter which is dependent on the voltage of the reader resonator. Having defined the state vector $\Phi = (U_r,U_s,U_r',U_s')$, one can further rewrite \eqref{eq:system_equation_differential} in terms of the Liouvillian formalism as
\begin{equation}
  \frac{\text{d} \Phi}{\text{d} \tau} = \mathcal{L} \Phi,
\end{equation}
where
\begin{equation} \label{eq:Liouvillian}
  \mathcal{L} =
  \begin{pmatrix}
    0                               & 0                           & 1        & 0                      \\
    0                               & 0                           & 0        & 1                      \\
    -\frac{1}{\barkappa}           & \frac{\kappa}{\barkappa}   & g_r(U_r) & 0                      \\
    \frac{\kappa}{\eta\barkappa} & -\frac{1}{\eta\barkappa} & 0        & -\frac{\gamma_s}{\eta} \\
  \end{pmatrix}.
\end{equation}
It is noted that the corresponding effective Hamiltonian $H_\text{eff} = \text{i} \mathcal{L}$ is PT-symmetric when $g_r = \gamma_s$ and $\eta = 1$, as implicated in \figref{fig:figure01}(b).

Figs. \ref{fig:figure01}(c) and \ref{fig:figure01}(d) plot the eigenfrequency evolution of an ideal unperturbed linear PT-symmetric electronic dimer when $\gamma = g_r = \gamma_s$, $\eta = 1$ and $\kappa = 0.2$, which exhibits two parametric regions. Since two of the four eigenfrequencies are opposites of each other, only two of them with positive real part are illustrated here. When $\gamma < \gamma_\text{PT}$, the system exhibits four real eigenfrequencies and therefore is in PT-symmetric phase. As $\gamma$ is increased, a spontaneous symmetry breaking and a second-order degeneracy of eigenfrequencies occurs at the exceptional point where $\gamma_\text{PT} = \sqrt{2(1-\sqrt{\barkappa})/\barkappa}$. When $\gamma > \gamma_\text{PT}$, the system exhibits two pairs of complex conjugate eigenfrequencies and therefore is in symmetry-broken phase. The parameter $\gamma$ is the \emph{so-called} ``parameter of Hermiticity".

Our sensing principle is by monitoring the steady-state working frequency of the reader resonator in response to capacitance change of the sensor resonator. Therefore, the gain and loss are no longer balanced and the capacitance change is regarded as a perturbation to loss side. Unlike the eigenfrequency solving method proposed in Ref.~\onlinecite{schindler2011experimental} where $g_r = \gamma_s$, here we seek for solutions of the steady-state eigenfrequency with zero imaginary part together with the steady-state gain $g_{r,\infty}$.

In order to guarantee non-trivial solutions for the system equation, the determinant of the coefficient matrix in \eqref{eq:Liouvillian} should be zero. Therefore the characteristic equation $\text{det}(H_\text{eff}-\omega \mathbf{I}) = 0$ has to be satisfied for the eigenfrequency of the system, which yields
\begin{equation} \label{eq:char_eq}
  \begin{split}
    \omega^4 &- \frac{1 + \eta - \gamma_s g_r\barkappa}{\eta\barkappa}\omega^2 + \frac{1}{\eta\barkappa} \\
    &+ \text{i} \left( \frac{\eta g_r - \gamma_s}{\eta}\omega^3 - \frac{g_r - \gamma_s}{\eta\barkappa} \omega \right) = 0.
  \end{split}
\end{equation}
For any physical system, the angular frequency $\omega$ should be real. Therefore, the real part and imaginary part in \eqref{eq:char_eq} should respectively be zero, which yields
\begin{align}
    \omega^4     & = \frac{1+\eta-\gamma_s g_{r,\infty}\barkappa}{\eta\barkappa}\omega^2 - \frac{1}{\eta\barkappa}, \label{eq:char_eq_steady_a} \\
    g_{r,\infty} & = \gamma_s\frac{1-\barkappa\omega^2}{1-\eta\barkappa\omega^2}. \label{eq:char_eq_steady_b}
\end{align}
When $\eta \neq 1$, eliminating the steady-state gain $g_{r,\infty}$, one arrives at the characteristic equation $\omega^6 + c_1 \omega^4 + c_2 \omega^2 + c_3 = 0$, where $c_1 = \eta(\eta+2)\barkappa c_3- \gamma_s^2/\eta^2$ and $c_2 = -[1 + \eta(1+\barkappa)] c_3 - \gamma_s^2/\eta^2$ with $c_3 = -1/(\eta^2 \barkappa^2)$. As a result, the eigenfrequencies are analytically derived as
\begin{subequations} \label{eq:omegas}
  \begin{align}
    \omega_{1,4} & = \pm \sqrt{s+t-\frac{1}{3}c_1},                                                \\
    \omega_{2,5} & = \pm \sqrt{-\frac{1}{2}(s+t)+\text{i} \frac{\sqrt{3}}{2}(s-t)-\frac{1}{3}c_1}, \\
    \omega_{3,6} & = \pm \sqrt{-\frac{1}{2}(s+t)-\text{i} \frac{\sqrt{3}}{2}(s-t)-\frac{1}{3}c_1},
  \end{align}
\end{subequations}
where $s = \sqrt[3]{p+\sqrt{p^2+q^3}}$ and $t = \sqrt[3]{p-\sqrt{p^2+q^3}}$ with $p = -\frac{1}{27}c_1^3+\frac{1}{6}c_1c_2-\frac{1}{2}c_3$ and $q = -\frac{1}{9}c_1^2+\frac{1}{3}c_2$. It is evident that the eigenfrequencies are dependent of the coupling coefficient $\kappa$, loss parameter $\gamma_s$ and perturbation parameter $\eta$. When $\eta = 1$, \eqref{eq:char_eq_steady_b} degenerates into $g_{r,\infty} = \gamma_s$, in which case solving \eqref{eq:char_eq_steady_a} yields the same eigenfrequencies as the PT-symmetric case, \emph{i.e.}, $\omega_{1,4} = \pm \sqrt{\frac{\alpha + \sqrt{\beta}}{2\barkappa}}$ and $\omega_{2,5} = \pm \sqrt{\frac{\alpha - \sqrt{\beta}}{2\barkappa}}$, where $\alpha = 2-\gamma_s\barkappa$, $\beta = (2-\gamma_s^2\barkappa)^2-4\barkappa $. However, another two eigenfrequencies $\omega_{3,6} = \pm 1/\sqrt{\barkappa}$ can be obtained from the solutions of $1 - \barkappa\omega^2 = 0$, which corresponds to the saturation gain $g_{r,\infty}|_{\omega_{3,6}} = (1/\barkappa-1)/\gamma_s$.

\begin{figure}[!ht]
  \centering
  \includegraphics[width=3.3in]{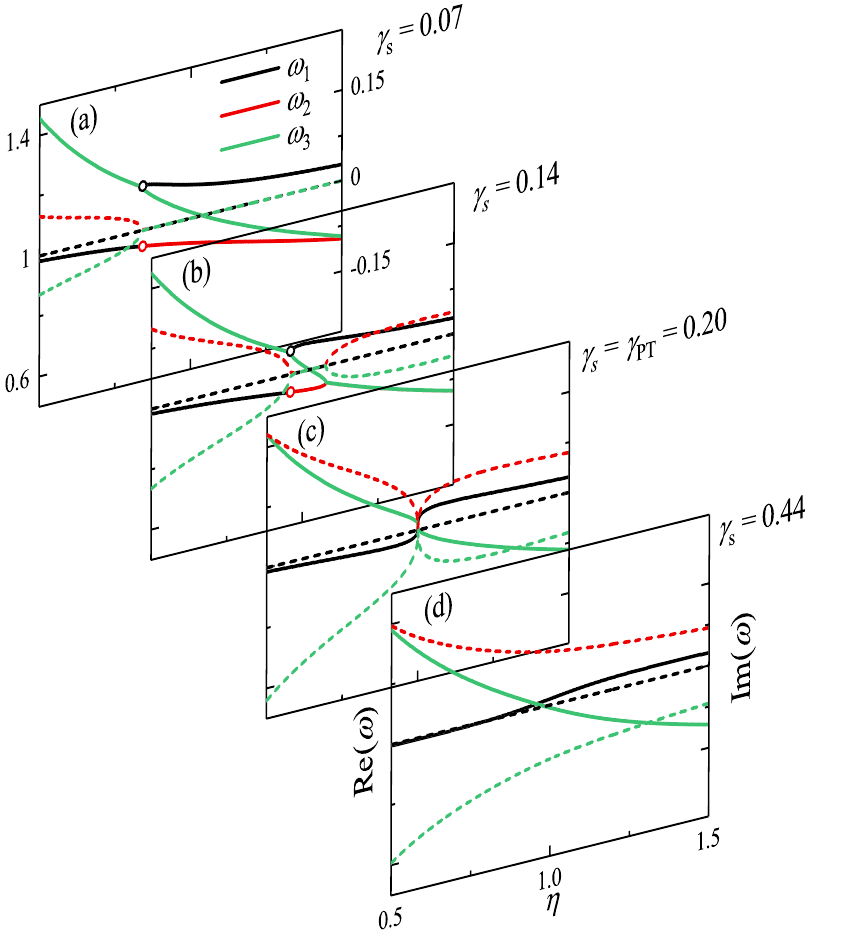}
  \caption{(Color online) Real part(solid lines) and imaginary part(dashed lines) of the eigenfrequency evolution versus $\eta$ when $\gamma_s$ is set to (a) 0.07, (b) 0.14, (c) 0.20 and (d) 0.44 with the same coupling parameter $\kappa=0.2$, which corresponds to the four labeled points in \figref{fig:figure01}(c). The steady-state frequency is one of the three eigenfrequencies with $\text{Im}(\omega)=0$ and requiring the least gain at the same time.}\label{fig:figure02}
\end{figure}

\figref{fig:figure02} illustrates the real and imaginary parts of the theoretical eigenfrequency evolution versus $\eta$ at different $\gamma_s$ when the coupling parameter $\kappa$ is fixed to 0.2. Here, the frequency $\omega$ is normalized by the natural frequency of the reader resonator $\omega_r = 1/\sqrt{L_r C_r}$. In the PT-symmetric phase of the linear unperturbed dimer where the system exhibits two real modes, the time-domain waveform of both gain and loss resonator has two frequency components corresponding to the two eigenfrequency branches when $\gamma<\gamma_\text{PT}$, as shown in \figref{fig:figure01}(c). However, for PT dimer with nonlinear gain implemented by active components, \emph{i.e.}, OP AMP, it becomes more involved. As the differential input voltage is increased, the output voltage can only increase to the nonlinear limit of the OP AMP. Beyond the linear range, the OP-AMP-based negative resistor yields a normal positive resistor. Due to the saturation effect of the OP AMP, the system cannot stabilize at the two eigenfrequencies at the same time but saturates to a steady state mode with the minimum saturation gain. As illustrated in \figref{fig:figure02}, three eigenmodes could exist, which implies the nonlinearity yields the increasing of the number of eigenmodes. Different modes will compete with each other and the mode requiring the least saturation gain will grow to reach the stable oscillation while suppressing other modes to acquire higher gain level \cite{assawaworrarit2017robust,kananian2020coupling}. Therefore, the conditions for mode $\omega_n$ to be the steady-state mode include: (1) $\omega_n$ is a real mode such that $\text{Im}(\omega_n) = 0$; (2) $g_{r,\infty}|_{\omega=\omega_n}$ is the minimum. We use eigenfrequency $\omega_n$ to represent the corresponding eigenmode hereafter.

When the perturbation parameter $\eta$ introduced by the sensor capacitance changes, the steady-state mode will change accordingly. For different perturbation starting points defined by $\gamma_s$, the steady-state mode will evolve in different ways. As shown in \figref{fig:figure02}(a) and (b), when the perturbation point is chosen in PT-symmetric phase (i.e., label 1 and 2 in \figref{fig:figure01}(c)), two exceptional points $\text{EP}_\eta$ ($\eta = \eta_{1,2}$) exsit in the eigenfrequency evolution with respect to $\eta$, which are solutions of the discriminant $\Delta(\eta) = p(\eta)^2 + q(\eta)^3 = 0$. For $\gamma_s = 0.07$ (or $\gamma_s = 0.14$), $\eta_{1,2}$ equals to 0.841 (0.961) and 1.525 (1.079) respectively. The system exhibits one real mode $\omega_1$ and two complex modes $\omega_2$ and $\omega_3$ with conjugate eigenfrequencies when $\eta < \eta_1$; three real modes when $\eta_1 < \eta < \eta_2$; one real mode $\omega_1$ and two complex modes $\omega_2$ and $\omega_3$ again when $\eta > \eta_2$. The first $\text{EP}_\eta$ when $\eta = \eta_1$ is where $\omega_2$ and $\omega_3$ degenerate from two complex modes to one real mode; and the second $\text{EP}_\eta$ when $\eta = \eta_2$ is where $\omega_2$ and $\omega_3$ degenerate from two real modes to one real mode. Therefore, when $\gamma_s < \gamma_{\text{PT}}$ there could be three real modes at the same time even in the absence of PT symmetry, implying that the system have a perturbed PT-symmetric phase with unbalanced gain and loss.

\begin{figure}[!ht]
  \centering
  \includegraphics[width=3.2in]{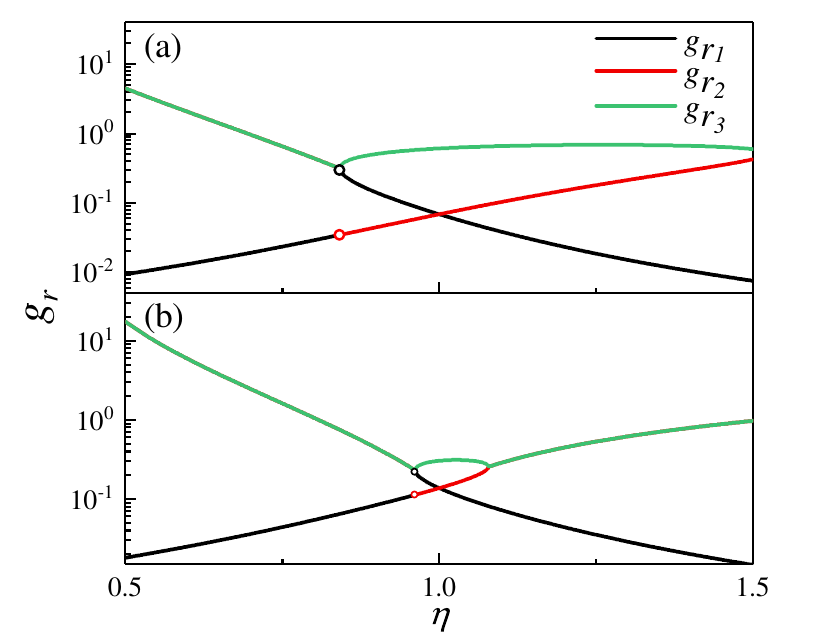}
  \caption{(Color online) Saturation gain required by each mode when (a) $\gamma_s=0.07$ and (b) $\gamma_s=0.14$ according to \eqref{eq:char_eq_steady_b}. \label{fig:figure03}}
\end{figure}

In order to determine the steady-state mode, the saturation gain required by each mode is evaluated by substituting the calculated eigenfrequencies \eqref{eq:omegas} to \eqref{eq:char_eq_steady_b}. Figs. \ref{fig:figure03}(a) and \ref{fig:figure03}(b) display the saturation gain evolution with respect to $\eta$ when $\gamma_s = 0.07$ and $\gamma_s = 0.14$, respectively. Now, the steady-state frequency can be determined according to Figs. \ref{fig:figure02}(a), \ref{fig:figure02}(b) and Figs. \ref{fig:figure03}(a), \ref{fig:figure03}(b). It is shown that for the perturbation point when $\gamma_s < \gamma_\text{PT}$, the steady-state frequency corresponds to the red solid line when $\eta \leq \eta_1$, the blue solid line when $\eta_1 < \eta \leq 1$ and the red solid line again when $\eta \geq 1$. Therefore, the system would operate in the lower frequency branch when $ \eta = C_s/C_r < 1$ and in the higher frequency branch when $\eta > 1$, with the working frequency being decreased as $C_s$ is increased.

As for $\gamma_s = \gamma_\text{PT}$ and $\gamma_s=0.44$ depicted in Figs. \ref{fig:figure02}(c) and \ref{fig:figure02}(d), only one real mode, $\omega_1$, is the steady-state mode since no mode competition occurs. When the perturbation point is chosen at EP (label 3 in \figref{fig:figure01}(c)), the coalesced eigenmodes will bifurcate into three different eigenmodes, one of which exhibits real eigenfrequencies while the other two modes exhibit complex conjugate eigenfrequencies as indicated by \figref{fig:figure02}(c). When perturbing from a point in the symmetry-broken phase (label 4 in \figref{fig:figure01}(c)), the previous two complex eigenmodes will bifurcate into one real mode and two complex modes. Comparing the four scenarios illustrated in \figref{fig:figure02}, the steady-state frequency changes most significantly when the perturbation point is chosen at EP and therefore has high sensitivity nearby the exceptional point. However, for wide range capacitive sensing, greater sensitivity occurs when perturbation point $\gamma_s$ is chosen in PT-symmetric phase; and the smaller the $\gamma_s$ is, the greater the sensitivity will be.

\begin{figure}[!ht]
  \centering
  \includegraphics[width=3.2in]{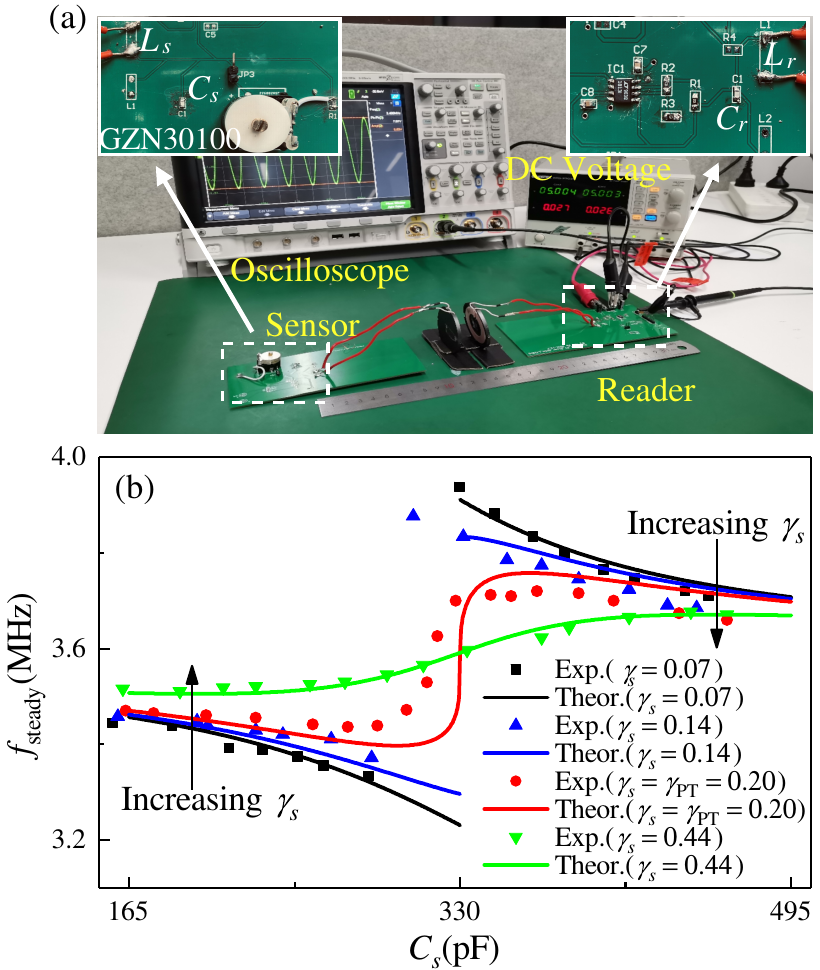}
  \caption{(Color online) (a) Experimental setup and (b) experimental results (markers) of steady state frequency evolution as $C_s$ changes in comparison with analytical results (solid lines) in \figref{fig:figure02} for various $\gamma_s$. \label{fig:figure04}}
\end{figure}

\paragraph*{Experimental verification \label{sec:experiment}}
In the experiment, see \figref{fig:figure04}(a) for the setup, a dual OP AMP chip LT1813 with $\pm5\rm{V}$ DC power supply was used on the reader side. One of the OP AMP is designed as the negative resistor while the other is functioned as a voltage buffer to avoid the influence of direct probe measurement. On the sensor side, a parallel passive RLC resonator with a GZN30100 trimmer capacitor is used to mimic the capacitive sensor. The reader and sensor coils are coaxially aligned, whose separation distance is fixed to $d = 1.8~\si{cm}$, yielding the coupling parameter $\kappa = 0.2$. The component values are: $C_r = 330~\si{pF}$, $L_r = L_s = 6.2~\si{\micro H}$, $R_f = 47.5~\si{\kilo\ohm}$ and $R_s = R_r = 1/\gamma_s \sqrt{L_r/C_r}$.

\figref{fig:figure04}(b) displays the experimental results of steady-state reader frequency as a function of the sensor capacitance $C_s$ (markers) in comparison with that obtained by the analytical derivation (solid lines) given in \figref{fig:figure02}, which shows a good agreement. It is shown that the proposed readout system can work in three different modes: two-branch mode, EP mode and symmetry-broken mode. For $\gamma_s < \gamma_\text{PT}$ (see the black and blue lines), the system operates in the two-branch mode, \emph{i.e.} at the lower and higher frequency branches when $C_s<C_r$ and $C_s>C_r$ respectively, with $C_s = C_r$ being a discontinuity point. In addition, on each frequency branch, the frequency is decreased as $C_s$ is increased with the sensitivity becoming larger as $C_s$ approaches $C_r$. This mode is more suitable for wide-range wireless capacitance readout. In EP mode when $\gamma_s = \gamma_\text{PT}$ (red line), the frequency evolution shows similar characteristic as two-branch mode except for nearby EP where a dramatic sensitivity occurs. Therefore, the proposed system can detect small capacitance changes by setting the system working nearby EP. In symmetry-broken mode when $\gamma_s > \gamma_\text{PT}$ (green line), the working frequency increases continuously as $C_s$ is increased. As a result, the proposed readout mechanism enables point-to-point wireless capacitance sensing under different modes by changing the perturbation point $\gamma_s$.

\paragraph*{Conclusions \label{sec:concl}}
In summary, we have demonstrated that PT-symmetric electronic dimer provides an real-time wireless sensing scheme apart from measuring the dip frequency shift in the reflectance spectrum through a VNA. Thanks to the nonlinear saturation effect of OP AMPs, the proposed system can realize a VNA-free and real-time wireless capacitance readout by directly monitoring the steady-state working frequency of the reader resonator. Since the capacitance change of the sensor resonator is regarded as a perturbation to the PT-symmetric system, no manual tuning of the reader capacitor is needed; therefore, the system is more suitable for sensing applications outside the laboratory. We also expect that for such steady-state-frequency-based system a higher sensitivity can be obtained by using PT-symmetric circuit with higher order EP, which will be detailed analyzed in a future work.

\vspace{-.5em}
\section*{Acknowledgements \label{sec:ack}}\vspace{-1em}
T.D. acknowledges support from National Natural Science Foundation of China (NSFC) under grant no. 51977165.

\vspace{-.5em}
\section*{Conflict of Interest \label{sec:interestsDeclartion}}\vspace{-1em}
The authors have submitted a patent application related to the presented work.

\vspace{-.5em}
\section*{Data Availability \label{sec:dataAvailability}}\vspace{-1em}
The data that support the findings of this study are available from the corresponding author upon reasonable request.

\bibliography{main} 
\onecolumngrid      

\end{document}